# Multi-shot Echo Planar Imaging for accelerated Cartesian MR Fingerprinting: an alternative to conventional spiral MR Fingerprinting

Arnold Julian Vinoj Benjamin[a,b,*], Pedro A. Gómez[c,d], Mohammad Golbabaee[a], Zaid Mahbub[b], Tim Sprenger[d], Marion I. Menzel[d] Michael Davies[a] and Ian Marshall[b]

[a]School of Engineering, Institute for Digital Communications, University of Edinburgh, United Kingdom; [b]Centre for Clinical Brain Sciences, University of Edinburgh, United Kingdom; [c]Computer Science, Technische Universität München, Munich, Germany; [d]GE Global Research, Munich, Germany

*Corresponding Author at: School of Engineering, Institute for Digital Communications, University of Edinburgh, United Kingdom

## Conflicts of Interest

## Acknowledgements

The research leading to these results has received funding from the European Union's H2020 Framework Programme (H2020- MSCA-ITN-2014) under grant agreement no 642685 MacSeNet, the Engineering and Physical Sciences Research Council (EPSRC) platform grant, number EP/M019802/1 and the Scottish Research Partnership in Engineering (SRPe) award, number SRPe PECRE1718/17.

# ABSTRACT

**Purpose:**

To develop an accelerated Cartesian MRF implementation using a multi-shot EPI sequence for rapid simultaneous quantification of T1 and T2 parameters.

**Methods:**

The proposed Cartesian MRF method involved the acquisition of highly subsampled MR images using a 16-shot EPI readout. A linearly varying flip angle train was used for rapid, simultaneous T1 and T2 quantification. The accuracy of parametric map estimations were improved by using an iterative projection algorithm. The results were compared to a conventional spiral MRF implementation. The acquisition time per slice was 8 s and this method was validated on a phantom and a healthy volunteer brain in vivo.

**Results:**

Joint T1 and T2 estimations using the 16-shot EPI readout are in good agreement with the spiral implementation using the same acquisition parameters (deviation less than 3% for T1 and less than 4% for T2) for the healthy volunteer brain. The T1 and T2 values also agree with the conventional values previously reported in the literature. The visual quality of the multi-parametric maps generated by the multi-shot EPI-MRF and spiral - MRF implementations were comparable.

**Conclusion:**

The multi-shot EPI-MRF method generated accurate quantitative multi-parametric maps similar to conventional Spiral - MRF. This multi-shot approach achieved considerable k-space subsampling and comparatively short TR's in a similar manner to spirals and therefore provides an alternative for performing MRF using an accelerated Cartesian readout; thereby increasing the potential usability of MRF.

*Abbreviations:* MRF, Magnetic Resonance Fingerprinting; EPI, echo planar imaging; TR, repetition time.

# 1. INTRODUCTION

Quantitative MRI (q-MRI) is fast emerging as an extremely useful modality in diagnostic MR imaging because these images provide clinicians additional information that helps in more accurate diagnosis, improved disease monitoring and better treatment planning [1, 2, 3]. Quantitative parameters like proton density (PD), T1 and T2 relaxation times, etc. vary for normal and abnormal tissues and can give an indication of neurodegenerative disorders in the brain that are not readily detectable from conventional structural MR images [4, 5, 6]. The estimation of tissue parameters helps in greater tissue discrimination, segmentation and classification to improve disease detection and monitoring. For example, T1 mapping has various applications such as the detection of neurodegenerative disorders like multiple sclerosis (MS) [7], Alzheimer's disease [8], assessment of myocardial infarction [9] and characterizing fiber bundle anatomy in diffusion MRI [10] while T2 mapping is used for applications in ageing and cognitive decline [11], quantification of myocardial edema [12] and evaluation of articular cartilage damage in the knee [13, 14]. However, clinical time constraints have prevented the widespread clinical use of parametric mapping techniques [15, 16]. Recent emergence of rapid parametric mapping techniques such as Magnetic Resonance Fingerprinting (MRF) [17] and its various extensions [18, 19, 20] have shown that it is possible to generate multiple quantitative parametric maps simultaneously in a very short scan duration that is clinically feasible. MRF offers a completely different approach to quantification of images compared to conventional q- MRI because it involves the deliberate variation of acquisition parameters which generate a unique signal evolution (i.e. the signal evolutions are a function of acquisition parameters like flip angle (FA), repetition time (TR) etc.) for each tissue [20]. The unique signal evolutions or 'fingerprints' that depend on underlying tissue properties such as relaxation times are then matched to a precomputed dictionary to generate quantitative maps.

The first MRF implementation was able to simultaneously quantify T1, T2 and off-resonance effects and was based on a balanced steady state free precession (bSSFP) sequence which was sensitive to field inhomogeneities and produced banding artefacts [17]. These effects were mitigated by the use of an unbalanced steady state free precession sequence (SSFP) for multiparametric quantification [18, 21, 22, 23]. The most commonly used sampling strategy in MRF is interleaved spiral sampling because it allows considerable subsampling of k-space and also provides more control for efficient traversal of the k-space trajectory [17, 18]. Despite its numerous advantages, the spiral sampling scheme has been shown to be susceptible to gradient inaccuracies [24] and high frequency artefacts due to nonsampling of k-space corners [25] and its availability is limited which prevents its widespread use in clinical protocols [26].

Cartesian sampling schemes for MRF primarily based on single-shot Echo Planar Imaging (EPI) that have been proposed are promising but are not a like-for-like comparison with the spiral sampling strategy for MRF [27, 28, 29, 30, 31]. This is because single-shot EPI implementations do not allow subsampling in a similar manner to the spiral scheme and therefore the entire k-space has to be traversed for every frame during acquisition. This results in much longer TR's than would be achievable with spiral sampling and also places a burden on the gradient performance of the scanner due to the short echo spacing necessary to minimise image distortions in single-shot EPI [32]. Simulation results show that shorter TR's result in better T1 and T2 sensitivity for certain MRF sequences (see supplementary material).

In this study, a multi-shot EPI - MRF approach is proposed that not only allows considerable k-space subsampling but can also achieve shorter TR's that are comparable to conventional spiral MRF implementations in a sufficiently short scan duration. Multi-shot EPI can yield better SNR, reduced blurring and lower ghost intensity while it also reduces the burden on gradients and RF hardware such as gradient amplitude and slew rate compared to single-shot EPI [32]. It also has the advantage of reduced distortions due to magnetic field inhomogeneity [33]. Unlike spiral MRF, multi-shot EPI-MRF has a solid theoretical basis in terms of compressed sensing theory [34, 35]. We used an unbalanced steady state free precession ('unbalanced SSFP') sequence with a linearly varying flip angle (FA) ramp for the rapid generation of accurate quantitative multi-parametric maps at reduced number of repetitions [18]. An Iterative Projection Algorithm (IPA) called BLoch matching response recovery through Iterated Projection (BLIP) was used to improve the accuracy of the generated parametric maps [34, 35]. The use of the fast Fourier Transform (FFT) instead of the much slower non uniform fast Fourier Transform (NUFFT) coupled with the avoidance of high frequency artefacts that appear in spirals makes the multi-shot EPI - MRF approach a good alternative to Spiral - MRF and could potentially increase the utility and robustness of MRF. This study also makes a direct comparison of quantitative maps generated by EPI-MRF and Spiral-MRF for the first time without modifying the underlying pulse sequence.

## 2. METHODS

### 2.1. Pulse Sequence Design

The original MRF paper that was based on a bSSFP sequence was sensitive to banding artefacts [17]. In order to overcome this, Jiang et al. [18] suggested the use of an unbalanced SSFP sequence also sometimes called FISP sequence. In the multishot EPI - MRF method introduced here, we also used an unbalanced SSFP sequence but we made the following changes compared to previous papers: i) a variable flip angle ramp instead of pseudorandom FA's was used to improve the T1 and T2 quantification efficiency in fewer number of repetitions (N) [23]. Figure 1 shows the simultaneous T1-T2 sensitivity of exemplary values of grey matter (GM), white matter (WM) and cerebrospinal fluid (CSF) at 3T that were simulated for the unbalanced SSFP sequence using the Extended Phase Graph (EPG) model for a) Linear Ramp FA Pattern from 1° to 70° with N = 500 repetitions and b) Pseudorandom FA pattern with N = 1000 repetitions that was used by Jiang et al. [18]. It can be seen that by using the Linear Ramp FA pattern, a very similar T1-T2 sensitivity for GM and WM is achieved in only half the number of repetitions and a significantly better T1-T2 sensitivity for CSF can be achieved when compared to the pseudorandom FA pattern. ii) a subsampled Cartesian readout using 16-shot EPI was used to eliminate regridding, perform faster reconstruction and avoid high frequency artefacts that appeared in spiral readouts. The multi-shot EPI readout provides entire k-space coverage and is especially suited when iterative algorithms are used for reconstruction as the non-sampling of k-space corners in spirals gives rise to high frequency artefacts as shown by Cline et al. [25]. Moreover, EPI sequences have been used clinically for over 20 years and the artefacts that arise from EPI are better understood by clinicians when compared to spirals. Therefore, it has a great potential to be easily adopted in clinical protocols.

The acquisition consisted of a series of highly subsampled gradient echo images that were obtained using a 16-shot EPI readout (see Fig. 2 a). 8 unique lines of $k_y$-space data were acquired for each shot at a slightly different FA that linearly varied between 1° to 70°. Gradient spoiling was introduced by the spoiler gradient $G_z$ (see Fig. 1a) to make it an unbalanced SSFP acquisition. The zero order gradient moments for $G_x$ and $G_y$ were nulled to ensure constant magnetization for each shot throughout the acquisition (see Fig. 2b). The minimum achievable TR and TE were used to minimize scan duration.

## 2.2. Sequence Parameters

The scanning was performed on a 3T GE MR750w scanner with a 12 channel receive only head RF coil (GE Medical Systems, Milwaukee, WI). The study was approved by the local ethics committee. 16-shot EPI - MRF and Spiral - MRF datasets were acquired from a tube phantom (Diagnostic Sonar, Livingston, UK) consisting of 11 tubes with different T1 and T2 values and a healthy volunteer using a variable FA ramp [23]. A linear ramp FA variation from 1° to 70° was used for acquiring 500 repetitions (N=500). The TR was set to 16 ms for both EPI - MRF and Spiral - MRF acquisitions to compare the two MRF implementations that were acquired using the variable FA ramp. Both acquisitions had bandwidth (BW) = 250 kHz, Field of View (FOV) = 22.5 x 22.5 cm2, 128 x 128 matrix size, 5 mm slice thickness and Inversion Time (TI) = 18 ms. The echo time (TE) was 2 ms and 3.5 ms respectively for the spiral and EPI acquisitions. The acquisition time for a single slice was 8 s. A reference scan with the EPI blips turned o was performed for the multi-shot EPI-MRF acquisition for phase correction of EPI raw data. In addition, a spiral-MRF dataset with pseudorandom FA train, varying TR and N = 1000 repetitions [18] was also collected for comparison with an established MRF method and to evaluate the performance of the proposed method. The healthy volunteer scans were also compared with the conventional T1 and T2 values that have been previously reported in literature [36]. WM and GM regions were extracted from the healthy volunteer brain images to calculate the accuracy of T1 and T2 quantification. Segmentation was performed by thresholding using the Matlab Image Processing Toolbox.

The MRF dictionary of magnetic resonance responses was pre-computed off-line using a Matlab implementation of the EPG formalism [37]. The EPG model is an efficient computational tool for quantitative simulations of signals [38, 18, 21] obtained from various MRI pulse sequences and is also widely used for characterizing signal evolutions in sequences used for relaxometry (i.e. characterizing relaxation parameters) [39, 40, 41]. This model is used to numerically compute the dictionary for MRF sequences by effectively modelling the pathways that lead to the formation of echoes [23, 21, 41]. A high resolution dictionary having a total of 23866 dictionary atoms was used with T1 values ranging from 40 ms to 2 s in steps of 20 ms and 2 s to 6 s in steps of 100 ms. The T2 values ranged from 20 ms to 120 ms in steps of 1ms, 120ms to 200ms in steps of 2 ms and 200 ms to 600 ms in steps of 10 ms. The inner product of each of the dictionary atoms and the measured response for each pixel was first computed and the parametric values of the dictionary atom that had the maximum correlation with the measured response was assigned to each pixel.

The dictionary was computed in approximately 5 minutes on a typical laptop computer with standard specifications. Figure 3 illustrates the sensitivity of a subset of the dictionary elements. The T1 sensitivity (16 fingerprints of dictionary elements with varying T1 ranging from 100 ms to 700 ms in steps of 40 ms and constant T2 = 100 ms) and T2 sensitivity (17 fingerprints of dictionary elements with varying T2 ranging from 20 ms to 100 ms in steps of 5 ms and constant T1 = 1000 ms) of the sequence for discriminating dictionary elements using a linear ramp FA variation from 1° to 70° are shown for 500 frames. Figure 2a shows that the T1 sensitivity is high throughout the acquisition and is enhanced by the initial inversion pulse whereas Figure 2b shows that the T2 sensitivity occurs mostly at higher flip angles (> 20°). Therefore, higher flip angles are needed for efficient T2 discrimination.

The reconstruction was done entirely in Matlab using the code adapted from the works done by Ma et al. [17] and Davies et al. [34, 35]. Two classes of reconstruction are considered: an IPA reconstruction that included Singular Value Decomposition (SVD) Compression in the Time Domain [42, 25, 43, 44]; and Dictionary matching (DM) sometimes called Matched Filter reconstruction as proposed in the original work on MRF [17], which is equivalent to a single iteration of the IPA reconstruction. The IPA reconstruction is motivated by compressed sensing theory [45, 46, 47, 48] and is shown to be capable of removing aliasing artefacts (in the reconstructed images) resulting from severe EPI style k-space subsampling. In the first iteration of IPA, DM is performed on the highly subsampled combined 12 coil zero-filled (ZF) images that are back projected. The back projection includes combining multi-coil measurements, a 2D inverse FFT for each temporal slice and a linear temporal compression where the compression bases are pre-calculated using the dominant SVD components of the fingerprint dictionary. The temporal compression is performed to primarily reduce the reconstruction time. The 12 channel multi-coil data is combined using SENSE reconstruction using sensitivity maps that were computed from the acquired data [49]. Briefly, each iteration of IPA consists of:

$$X^{j+1} = \sigma_D(X^j - \mu A^H(A(X^j) - Y))$$

where $Y \in C^{m \times N}$ are the undersampled k-space measurements across N temporal repetitions and multiple coils, μ is the step size which is adaptively selected through line search [35], $X_j \in C^{n \times N}$ are the spatio-temporal reconstructed images at iteration 'j' and $D \in C^{d \times N}$ denotes the pre-computed dictionary with '$d$' atoms ($d$ = 23866 atoms in this case). The forward and backward operators $A$, $A^H$ model the multi-coil sensitivities and 2D Fourier Transforms for the acquired subsampled data. σD denotes the DM step that is used in [35, 25] consisting of i) a search over the normalized dictionary atoms to replace the temporal pixels of $X^{j+1}$ with the maximum correlated fingerprints and ii) proton density rescaling. Therefore, the first iteration of IPA can be interpreted as an application of DM with proton density regularization [34, 35]. SVD compression-decompression is also applied iteratively in IPA reconstruction to reduce the complexity of the reconstruction [25, 43]. The first 20 dominant singular values were enough to robustly compress the dictionary. The IPA reconstruction was allowed to converge through multiple iterations to improve the data fidelity (i.e. to reduce the relative error between the quantitative estimate and the MRF measurements) resulting in more accurate tissue parametric estimations.

## 3. RESULTS

Figure 4 shows the temporal signal curve of one representative voxel from a subsampled EPI-MRF image along with its matched dictionary entry for a) phantom and b) healthy volunteer. Although both the phantom and healthy volunteer images are dominated by subsampling artefacts, the DM algorithm is still able to find the corresponding dictionary entry that has the maximum correlation with the acquired data showing its robustness to undersampling artefacts. Note that undersampling artefacts are regular due to uniform subsampling but the signal along the temporal domain is still noise-like which is similar to the Spiral-MRF case as shown by Jiang et al. [18]. This noiselike behaviour of the signal in the temporal domain facilitates effective discrimination between dictionary elements resulting in an accurate dictionary match. Figure 5 shows the highly aliased ZF images along with the unaliased IPA reconstructed images of the tube phantom and healthy volunteer at different repetition indexes 't' . It can be seen from the IPA reconstructed images that at lower repetition indexes (i.e. t = 44, 60) the images are predominantly T1 weighted. At higher repetition indexes (i.e. t = 350, 420) the images are more T2-weighted which is in agreement with the parameter encoding (i.e. FA train) used during acquisition. The signal intensity gradually increases due to the linear increase in the flip angles. Figure 6 shows the comparison of T1 and T2 maps of the tube phantom that were generated after DM (i.e. single iteration of IPA) for i) Spiral - MRF and ii) multi-shot EPI - MRF. The parametric maps are visually comparable for the two methods. Figure 7a shows the arrangement of tubes with different T1 and T2 values in the phantom while Figures 7b and 7c show the mean (± standard deviation) T1 and T2 values of each tube in the phantom for three different MRF acquisitions, namely i) Spiral - MRF (pseudorandom FA, varying TR and N = 1000 repetitions) in blue; ii) Spiral - MRF (ramped FA, TR = 16 ms and N = 500 repetitions) in orange and iii) EPI - MRF (ramped FA, TR = 16 ms and N = 500 repetitions) in grey. The mean T1 values are in close agreement for all three methods (less than 5% variation) while there are subtle differences in the mean T2 values (~ 10%) for the tube phantom. Figure 8 shows the generated multi-parametric maps of a healthy volunteer brain after the application of DM for i) Spiral - MRF and ii) EPI - MRF. Detailed structures can be clearly seen in the parametric maps of the healthy volunteer brain in both methods. Figures 9a and 9b show the segmentation of white matter (WM) and grey matter (GM) for the healthy volunteer comparing EPI - MRF and Spiral - MRF methods. The mean values and standard deviations of WM and GM shown in Figures 9c and 9d are in agreement with each other for both methods (i.e. less than 3% variation for T1 and less than 4% variation for T2). These values are also compared with literature values of Spiral - MRF (pseudorandom FA, varying TR and N = 1000 repetitions) method [18] and gold standard methods for conventional T1 and T2 measurements [36] as shown in Figures 9c and 9d. Figures 10 and 11 show the improved T1 estimations of EPI - MRF and Spiral - MRF respectively for a healthy volunteer after the application of the IPA reconstruction.

The IPA reconstruction improves the multi-parametric estimations through multiple iterations until the convergence of the algorithm. Also shown are the T1 maps generated after DM for comparison. The IPA reconstruction reduces the relative error between the quantitative estimate and the MRF measurements at every iteration and converges when this error becomes very small. The IPA reconstruction converged in 22 seconds (4 iterations) for EPI - MRF and 482 seconds (30 iterations) for Spiral - MRF when data from all coils ($N_c$ = 12) were used for reconstruction. A five-fold reduction (for EPI - MRF) and an eight-fold reduction (for Spiral - MRF) in the relative error was observed after the convergence of IPA reconstruction. Iterative reconstruction is particularly beneficial when we have limited data and its benefits are further highlighted in the supplementary material in a reduced coil scenario.

## 4. DISCUSSION

In this study, a new MRF scheme based on a vastly subsampled Cartesian readout that utilizes multi-shot EPI (i.e. EPI - MRF with ramped FA) has been introduced. Good quantification of T1 and T2 maps have been achieved both in phantom and healthy volunteer scans in about 8 s for the range of T1 and T2 values that normally occur in a human brain. The generated parametric maps of the proposed EPI - MRF method have been compared and are shown to be in good agreement with Spiral - MRF; thereby demonstrating the potential of Cartesian MRF as a suitable alternative to Spiral - MRF.

On comparison of the T1 maps of the tube phantom for EPI - MRF and Spiral - MRF in Figure 6, it can be seen that the T1 maps are visually comparable and the mean T1 values (see Fig. 7b) for each of the 11 tubes in the phantom are very close to each other with less than 3% deviation. This is due to the high T1 sensitivity of the encoding scheme used for the acquisition resulting in a good T1 quantification. However, the mean T2 values of Spiral - MRF and EPI - MRF show differences up to 10% when compared to the established Spiral - MRF (with pseudorandom FA, varying TR and N = 1000 repetitions) method [18] and can be seen from Figure 7c. The TR was set to 16 ms for both EPI - MRF and Spiral - MRF acquisitions for fair comparison even though the TR can be reduced to a minimum of 8 ms for the case of Spiral - MRF. However, there is a discrepancy between the estimated T2 values for different TR's with longer TR's resulting in higher estimates suggesting that the idealized EPG model used for the dictionary may have some inconsistencies and it merits further research. The variation in T2 may be caused by the encoding scheme which is comparatively less sensitive to T2 variations than T1 and can be seen from Figure 3 (i.e. the T1 sensitivity occurs throughout the acquisition whereas T2 sensitivity occurs only at flip angles greater than 20 degrees). Further optimization of the encoding scheme by the use of an optimized FA train instead of a linear ramp may yield more accurate T2 values. It may also be possible that the highest T2 sensitivity may occur only with a few FA's rather than a range of FA's (i.e. 1° to 70°) used in this acquisition. Therefore, constraining the acquisition to only a few FA's might provide better T2 sensitivity. Some preliminary work has already been done in this area [31].

The mean T1 values of white matter (WM) and grey matter (GM) for a representative healthy volunteer brain shown in Figure 9c for Spiral - MRF and EPI - MRF are almost identical (less than 3% deviation) and this demonstrates good T1 quantification for the healthy volunteer brain. In addition, the mean T2 values of both methods for WM and GM in the healthy volunteer (Fig. 9d) also correspond to each other with less than 4% variation. This shows that the T2 sensitivity of Spiral - MRF and EPI - MRF is very good for the range of T2 values that is normally present in a healthy volunteer brain at 3.0 T (i.e. 60 - 150 ms). Further validation is done by comparing the mean T1 and T2 values of WM and GM with the established Spiral - MRF (pseudorandom FA, varying TR and N = 1000 repetitions) technique [18] and also with previously reported T1 and T2 literature values of healthy volunteer human brain [36]. The mean T1 and T2 values of WM and GM are very close to the values obtained by Spiral - MRF (pseudorandom FA, varying TR and N = 1000 repetitions) [18] with less than 3% variation for T1 and less than 4% variation for T2 respectively (see Figs. 9c and 9d). The T1 and T2 values are also in agreement with gold standard T1 and T2 values of a representative healthy volunteer that has been reported in the literature [36]. This shows that the proposed multi-shot Cartesian EPI - MRF approach can generate similar T1 and T2 maps and can be a good alternative to the Spiral - MRF implementation.

On the other hand, there are some underlying drawbacks of T1 and T2 quantification through the MRF framework that also extend to the proposed EPI - MRF approach.

The quantification is not accurate in MRF when T1 and T2 values are very high (i.e. T1 > 2500 ms and T2 > 400 ms) due to the difficulty of discriminating dictionary entries at these values and this can be seen from simulations of dictionary atoms with high T1 and T2 values [23]. This can also be seen from the underestimation of T2 cerebrospinal fluid (CSF) values in both EPI - MRF and Spiral - MRF in Figure 8. In addition to that, EPI sequences are highly sensitive to B0 inhomogeneity caused by magnetic susceptibility variations [50, 33]. The B0 inhomogeneity is due to local differences in magnetic susceptibility that are particularly high at the interface between the glass tubes that hold the gadolinium doped agarose gels in the phantom resulting in higher mean T2 values in Figure 7c. The magnetic susceptibility differences are not as high in the human brain compared to the phantom resulting in better T2 estimation. However, there are slight ghosting artefacts in the T2 maps in Figure 8 which are EPI artefacts which might be caused by CSF pulsation disrupting the echo train and imperfect phase correction potentially causing phase errors [51]. They can be reduced by using Echo Time Shifting (ETS) that improves the phase error function in multi-shot EPI [52]. Due to the high T1 sensitivity of the acquisition, these artefacts are suppressed in the T1 maps but they affect the T2 maps. By enhancing the T2 sensitivity during the acquisition using an optimized FA train, these ghosting artefacts can be potentially reduced. By accurately correcting for magnetic susceptibility variations in the reconstruction and by using ETS to correct for phase errors, the proposed EPIMRF method would become more robust and this would the focus of future work.

Figure 10b shows there is a reduction in the relative error for EPI-MRF and Spiral MRF at each iteration until the convergence of the IPA algorithm. The relative error shows an approximate 5 times decrease for EPI - MRF and 8 times decrease for Spiral - MRF. The IPA reconstructions of multi-shot EPIMRF converge very quickly (in about 4 iterations) compared to the Spiral - MRF implementation (30 iterations) and could therefore result in a very fast implementation on the scanner. From Figure 10a, it can be seen that there are no artefacts introduced in the T1 maps for EPI - MRF after the application of IPA reconstruction whereas high frequency artefacts appear in the Spiral - MRF case (see Fig. 11). Although there is reduction in the relative error at each iteration, the non-sampling of k-space corners gives rise to errors that appear in the images as high frequency artefacts and this has already been shown by Cline et al. [25]. This is a fundamental limitation of the spiral sampling strategy and is not algorithm related. Note that low pass filtering should be performed to remove the high frequency artefacts that appear in Spiral - MRF when iterative reconstruction is used [25]. Since EPI - MRF provides full k-space coverage, high frequency artefacts are not present. The sampling used in EPI - MRF is less prone to errors due to non-sampling of kspace regions. EPI - MRF sampling is therefore fundamentally more suited for iterative reconstructions.

The reconstruction times are heavily dependent on the SVD compression-decompression that is used when moving from kspace to image space and vice versa [43, 25]. Each iteration uses SVD compression in the backward operation and SVD decompression in the forward operation. This provides a considerable reduction in reconstruction time. The reconstruction time was increased from 22 seconds to 206 seconds for EPI - MRF and from 482 seconds to approximately 6 hours for Spiral - MRF when SVD compression-decompression was not used in the reconstruction. The reconstructions were performed on a normal laptop computer with standard specifications. The convergence of the Spiral - MRF implementation is usually slower (i.e. both in time and in the number of iterations required) when compared to EPI - MRF (see Fig. 10b). This appears to be due to the bad conditioning of spiral sampling problem and the need for re-gridding to reconstruct spiral data [25]. In addition, each iteration is more expensive because spiral sampling uses a costlier NUFFT compared to the FFT used in EPI. Therefore, SVD compression-decompression is highly beneficial especially for Spiral - MRF in order to speed up the reconstruction time. Further reductions in the computation time are possible using an adaptive iterative algorithm with fast nearest neighbour searches for the DM step in the reconstruction [53]. The fast convergence of EPI - MRF and its robustness to high frequency artefacts make it naturally suitable for iterative reconstructions.

## 5. CONCLUSION

The multi-shot EPI-MRF method introduced here can provide joint quantification of multi-parametric maps such as T1 and T2 with good accuracy in a very short scan duration that is similar to Spiral - MRF. This multi-shot approach not only allows considerable k-space subsampling like spirals but also can achieve a short TR that is comparable to Spiral - MRF. As a result, it can be a suitable alternative for performing MRF using an accelerated Cartesian readout; thereby increasing the potential usability of MRF.

**FIGURES**

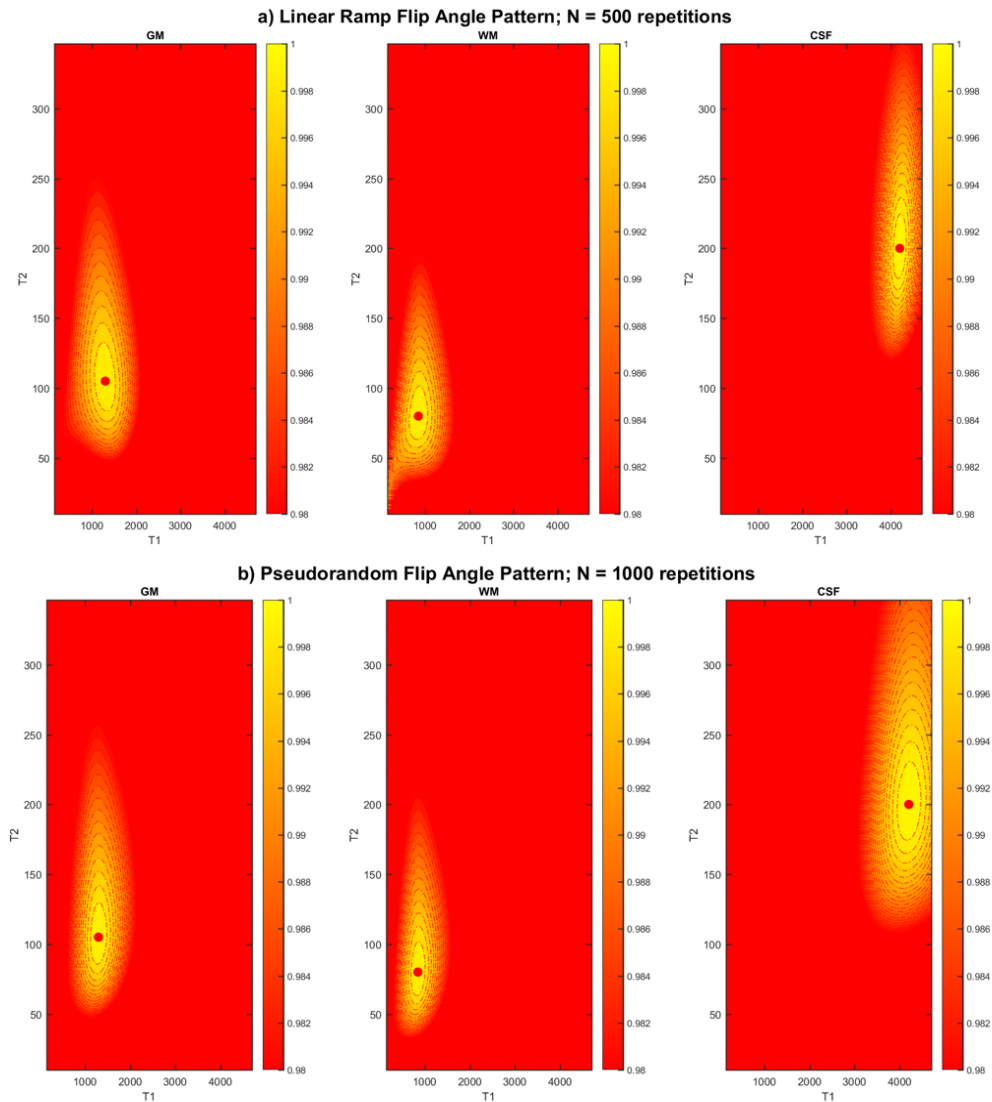

Figure 1: T1-T2 sensitivity of exemplary values of gray matter (GM), white matter (WM) and cerebrospinal fluid (CSF) at 3T that were simulated for the unbalanced SSFP sequence using the EPG model for a) Linear Ramp FA Pattern from 1° to 70° with N = 500 repetitions and b) Pseudorandom FA pattern with N = 1000 repetitions that was used by Jiang et al. [18].

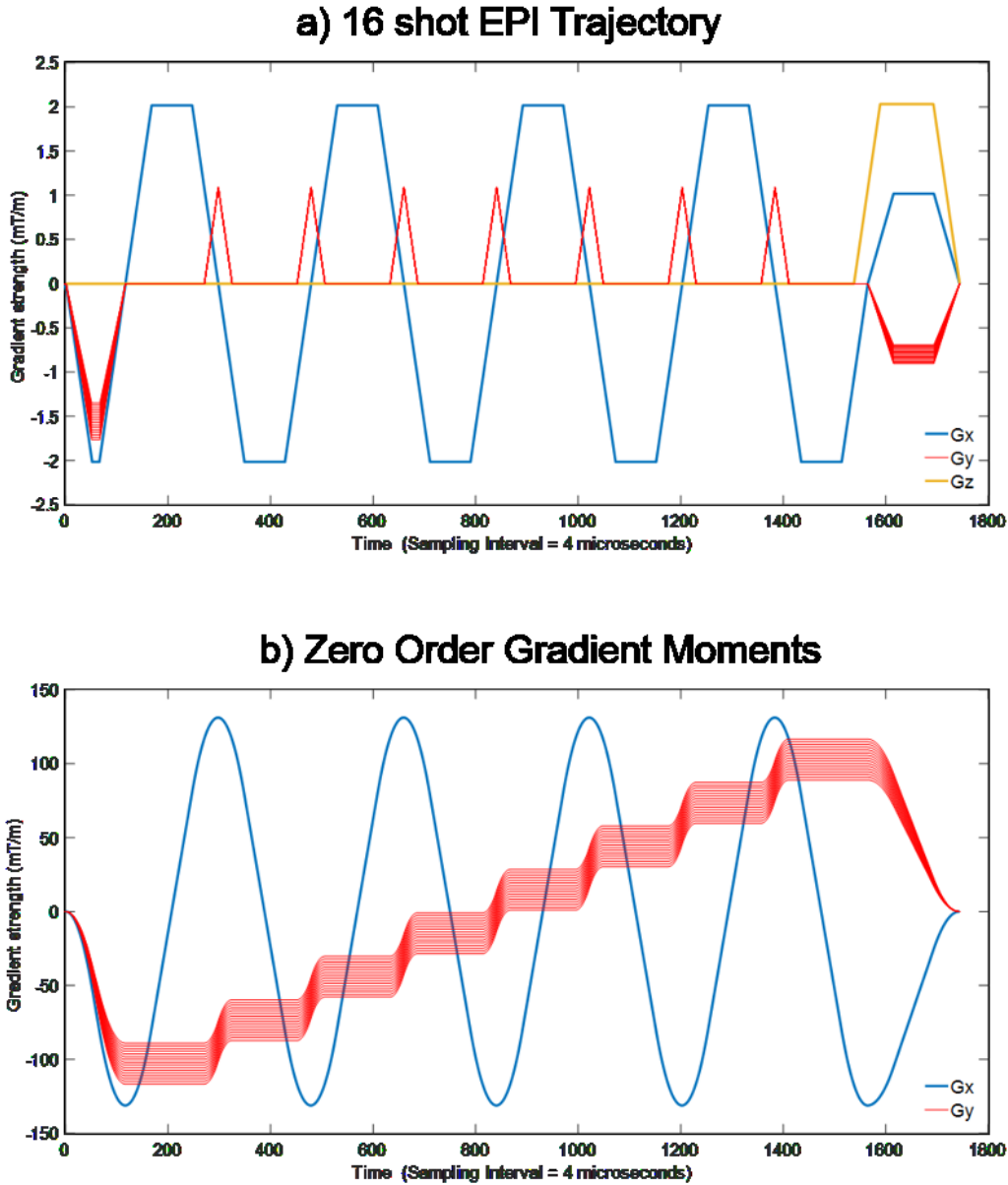

Figure 2: (a) The 16 shot EPI trajectory showing $G_x$, $G_y$ and $G_z$ gradients. Note that the $G_y$ gradients are slightly different for each of the 16 shots indicating that different lines of $k_y$ space are acquired at every shot. The spoiler gradient $G_z$ dephases the transverse magnetization for every TR making the sequence unbalanced [18]. (b) The corresponding x and y zero order gradient moments for $G_x$ and $G_y$ were nulled to ensure constant residual magnetization for each shot throughout the acquisition.

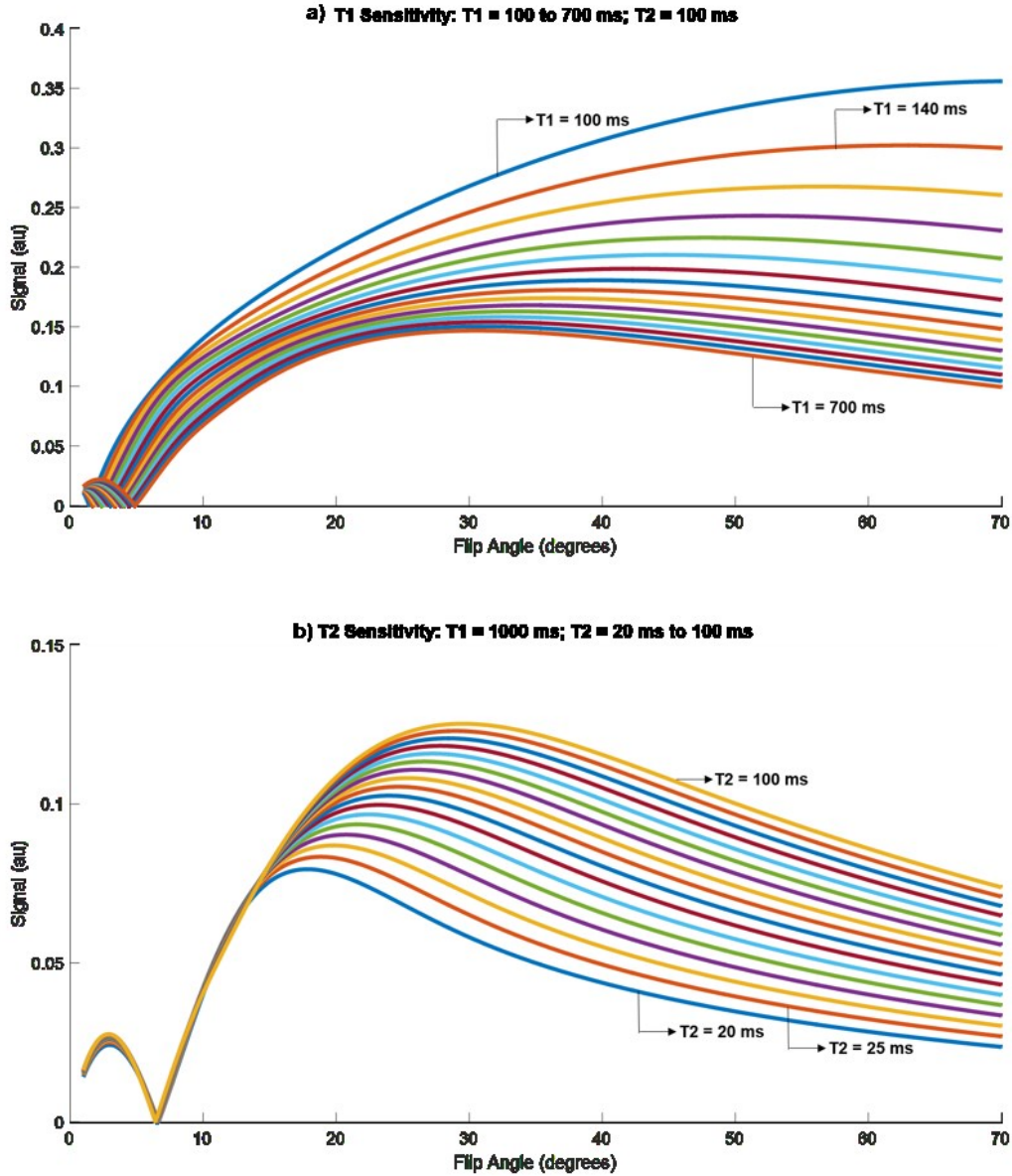

Figure 3: (a) Figure showing the T1 sensitivity and (b) T2 sensitivity of the sequence for discriminating dictionary atoms when a variable flip angle ramp that linearly varied between 1° to 70° was used during the acquisition for 500 repetitions. Note that the Inversion pulse causes the initial T1 discrimination in (a). These sensitivities were observed at practical T1 and T2 values. Only a subset of the high resolution dictionary is plotted for better visualization.

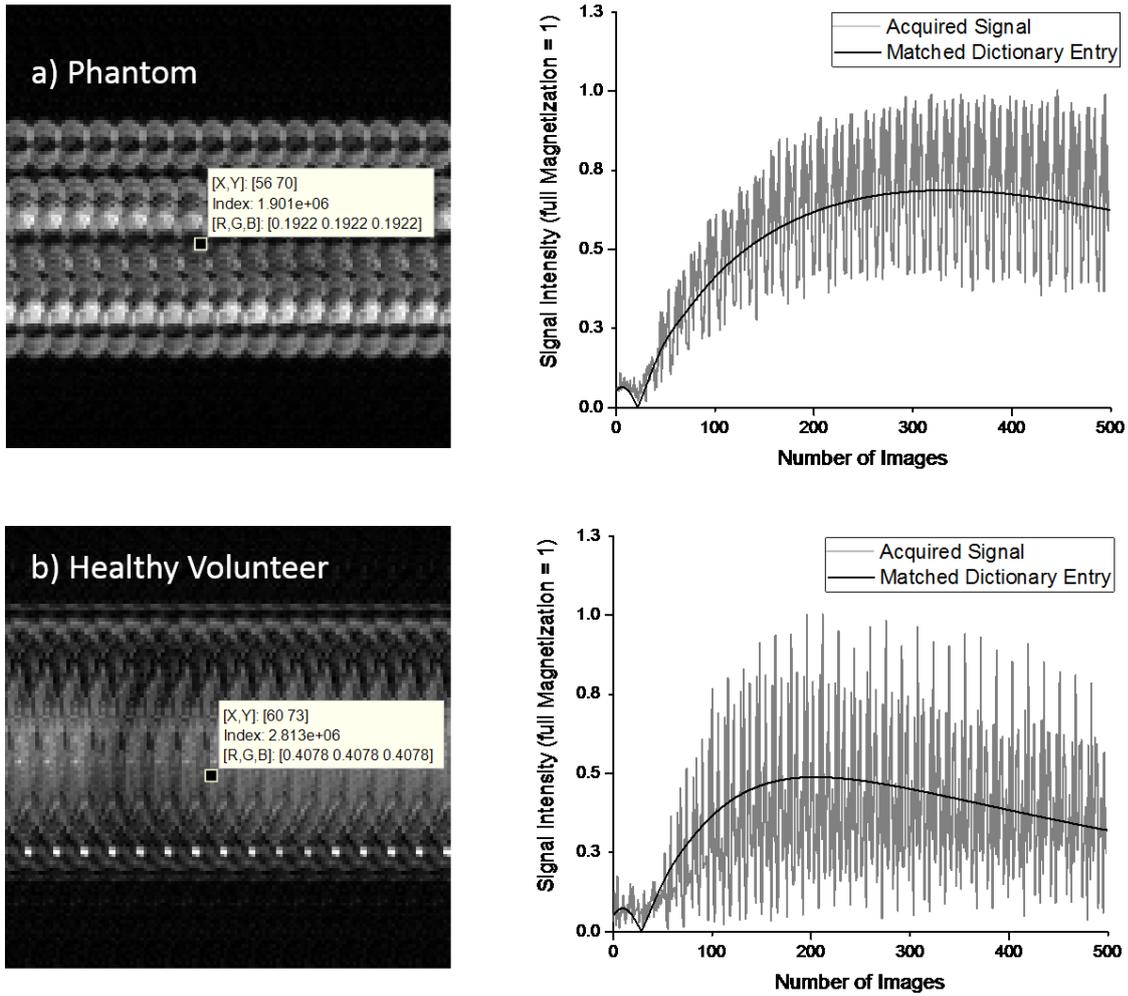

Figure 4: (a) Figure showing the temporal signal curve of one representative voxel from a subsampled EPI-MRF image along with its matched dictionary entry for a) phantom and b) healthy volunteer. Note that dictionary matching (DM) still works even in the presence of uniform subsampling artefacts in the image due to the noise-like behavior of the signal in the temporal domain.

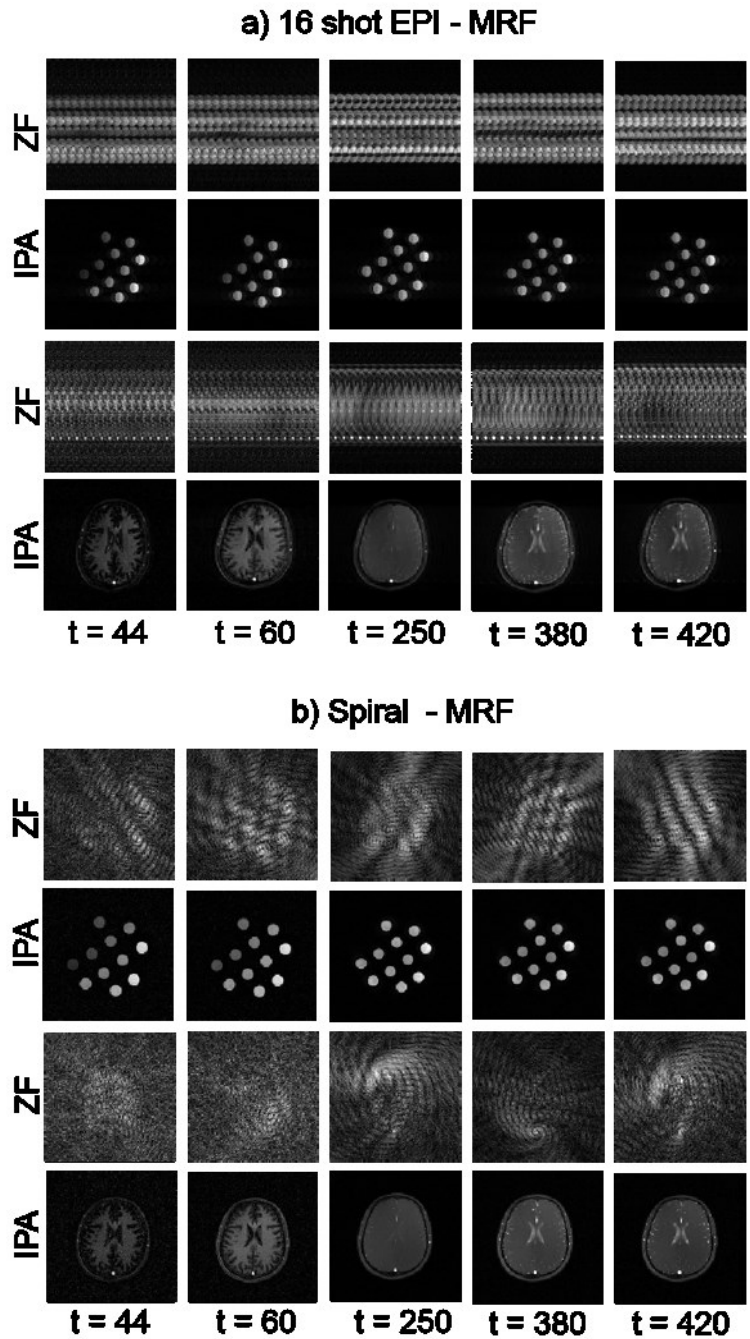

Figure 5: Figure showing the highly aliased zero-filled (ZF) images and Iterative Projection Algorithm (IPA) reconstructed images at different repetition indexes 't' of the tube phantom and the healthy volunteer for a) EPI - MRF (ramped FA, TR = 16 ms, N = 500 repetitions) and b) Spiral - MRF (ramped FA, TR = 16 ms, N = 500 repetitions).

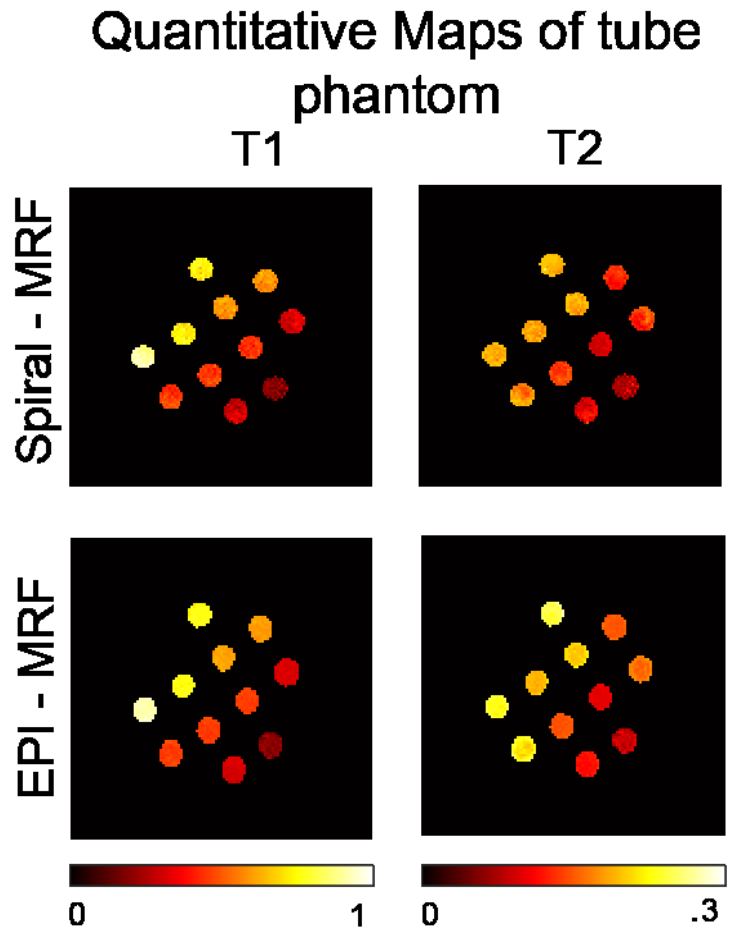

Figure 6: Figure showing the T1 and T2 maps (in seconds) of the tube phantom generated after Dictionary Matching (DM) for i) Spiral - MRF (ramped FA, TR = 16 ms, N = 500 repetitions) and ii) EPI - MRF (ramped FA, TR = 16 ms, N = 500 repetitions).

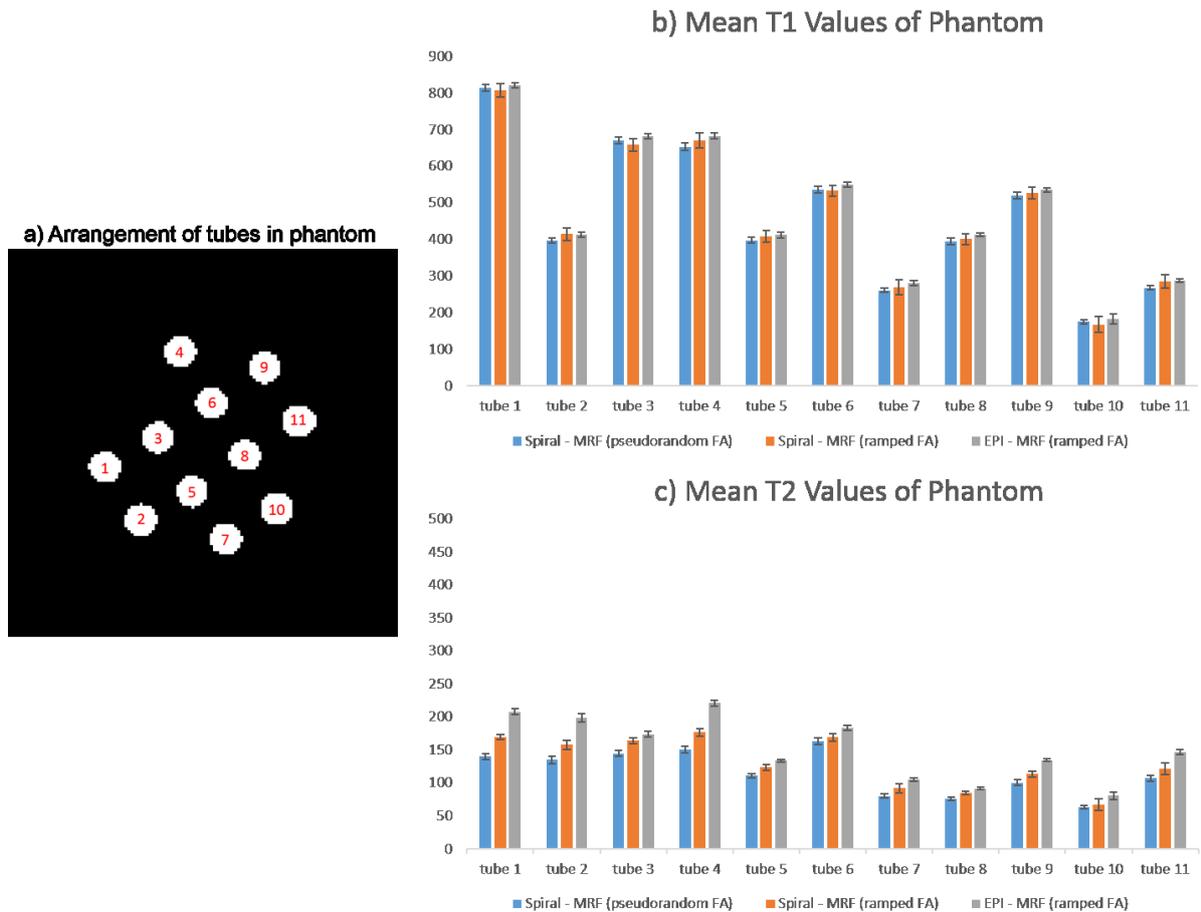

Figure 7: (a) Arrangement of tubes with different T1 and T2 values in phantom. (b) Mean T1 (± standard deviation) of various tubes in phantom comparing Spiral - MRF (pseudorandom FA, varying TR and N = 1000 repetitions) in blue, Spiral - MRF (ramped FA, TR = 16 ms and N = 500 repetitions) in orange and EPI - MRF (ramped FA, TR = 16 ms and N = 500 repetitions) in gray. (c) Corresponding mean T2 values (± standard deviation).

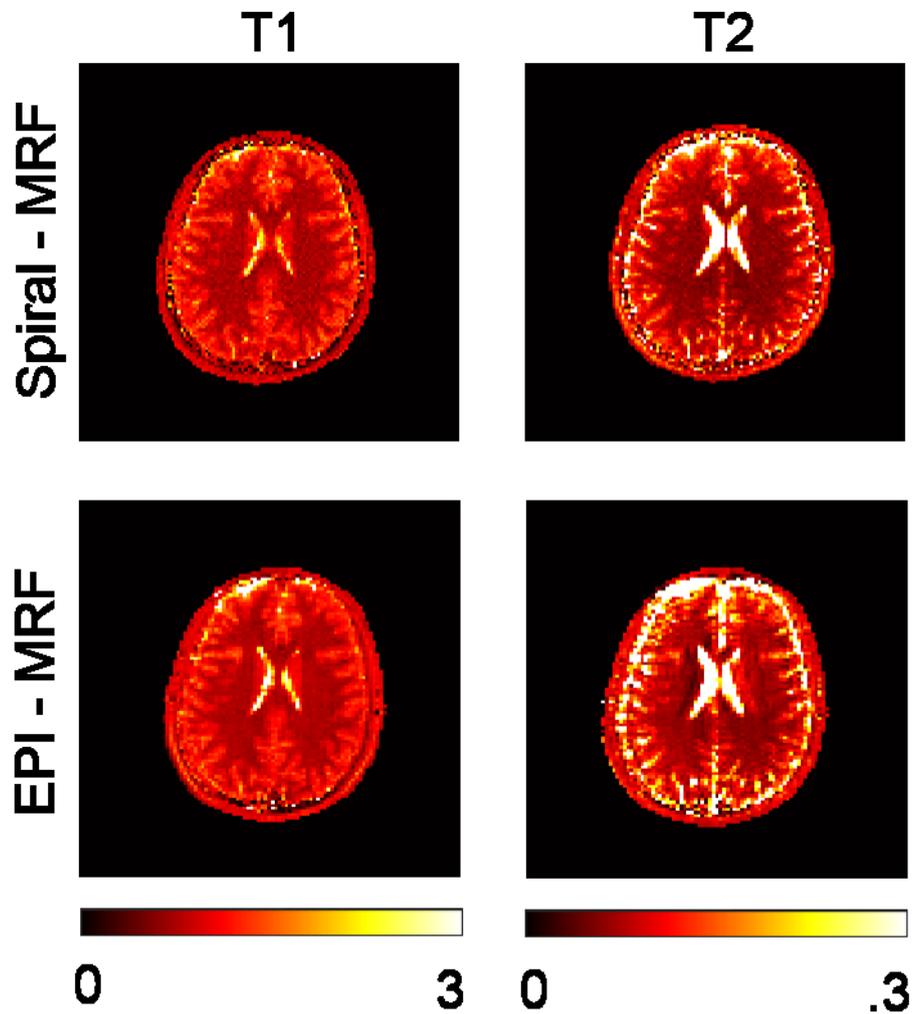

Figure 8: Figure showing the T1 and T2 maps (in seconds) of the healthy volunteer generated after Dictionary Matching (DM) for i) Spiral - MRF (ramped FA, TR = 16 ms, N = 500 repetitions) and ii) EPI - MRF (ramped FA, TR = 16 ms, N = 500 repetitions).

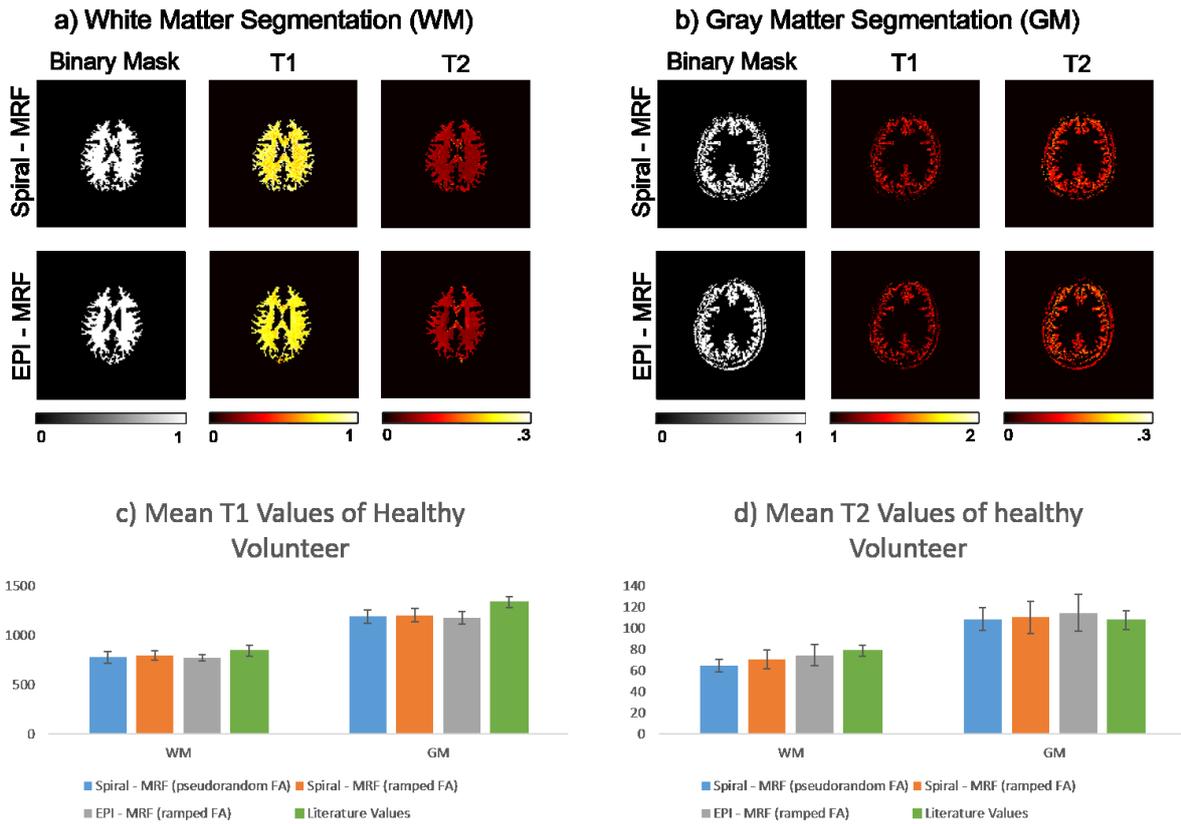

Figure 9: (a) Segmentation of white matter (WM) for the healthy volunteer showing the binary mask, T1 and T2 maps for Spiral - MRF and EPI - MRF. (b) Corresponding segmentation of gray matter (GM). (c) Mean T1 (± standard deviation) of WM and GM for healthy volunteer comparing Spiral - MRF (pseudorandom FA, varying TR and N = 1000 repetitions) in blue [18], Spiral - MRF (ramped FA, TR = 16 ms and N = 500 repetitions) in orange, EPI - MRF (ramped FA, TR = 16 ms and N = 500 repetitions) in gray and previously reported 'conventional literature values' in green [36]. (d) Corresponding mean T2 values (± standard deviation) of WM and GM.

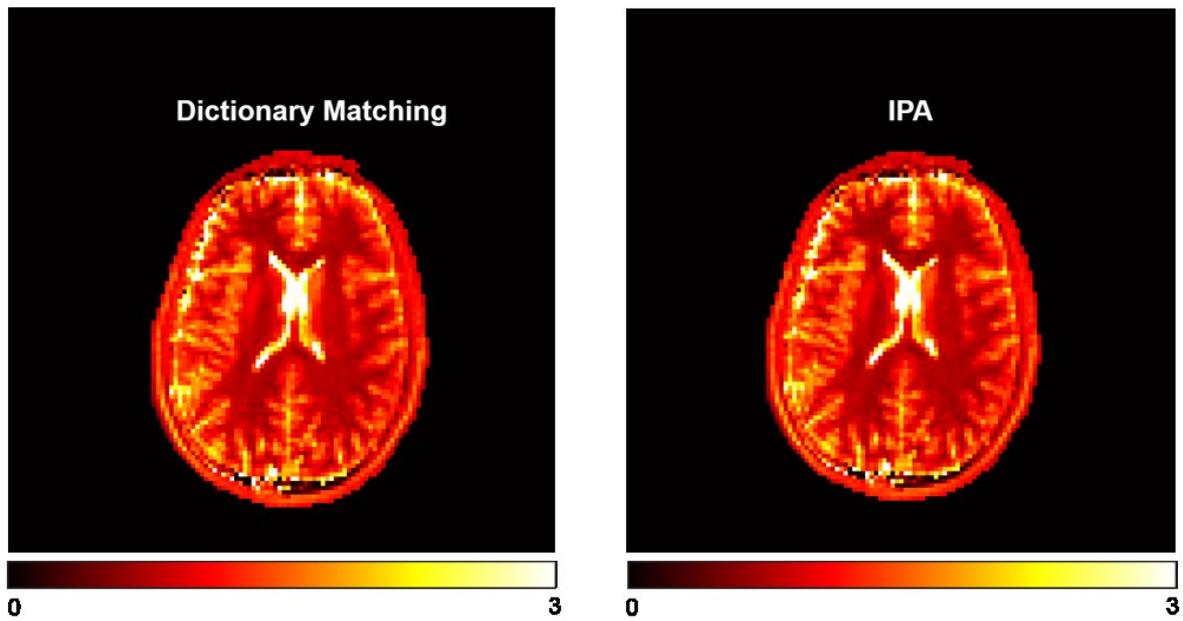

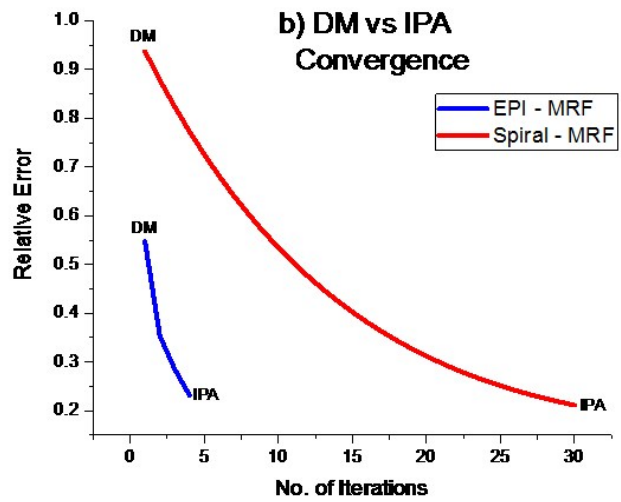

Figure 10: Figure showing the T1 maps of a healthy volunteer generated using Dictionary Matching (DM) and Iterative Projection Algorithm (IPA) respectively for EPI - MRF (a). A comparison of IPA convergence is shown for EPI - MRF and Spiral - MRF (b). Note that DM is equivalent to a single iteration of IPA.

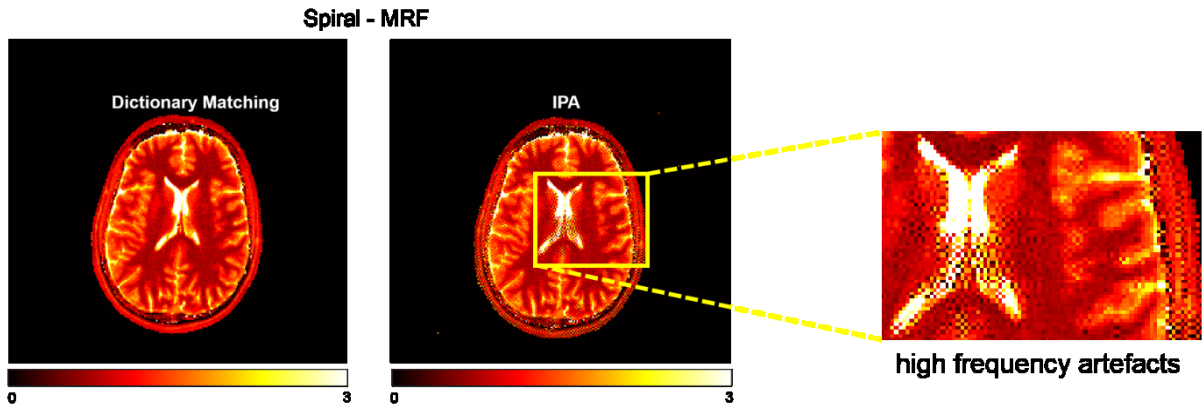

Figure 11: Figure showing the T1 maps of a healthy volunteer generated using Dictionary Matching (DM) and Iterative Projection Algorithm (IPA) respectively for Spiral - MRF. The enlarged image shows the appearance of high frequency artefacts after iterative reconstruction.

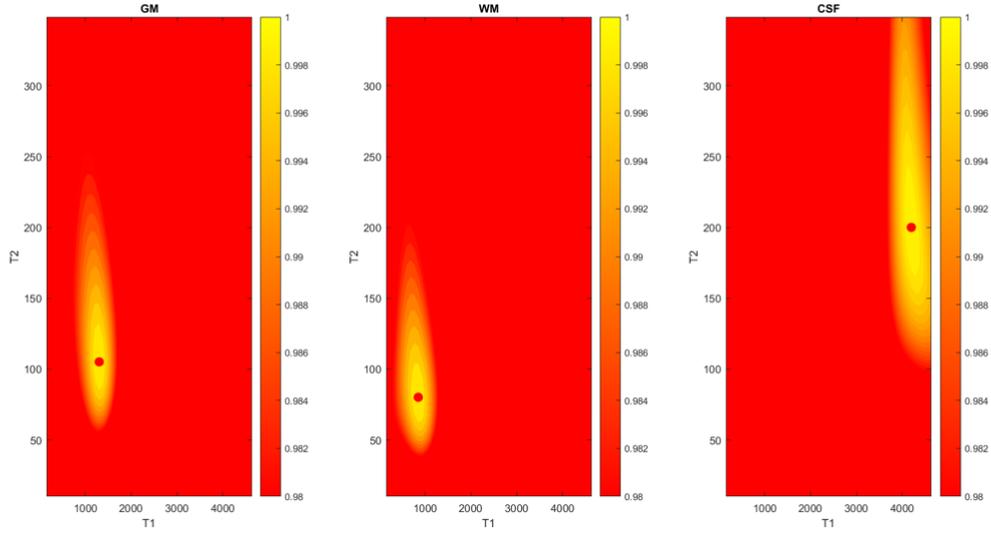

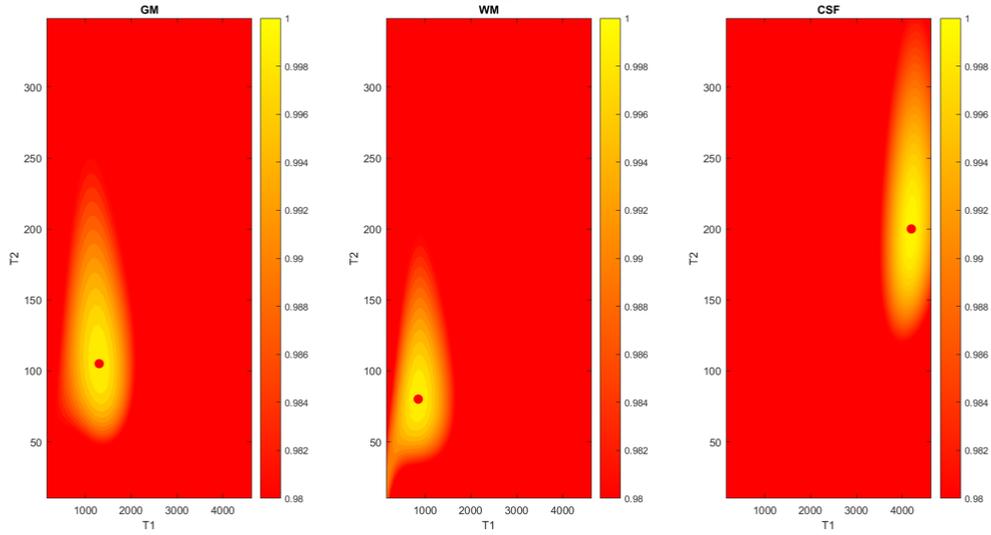

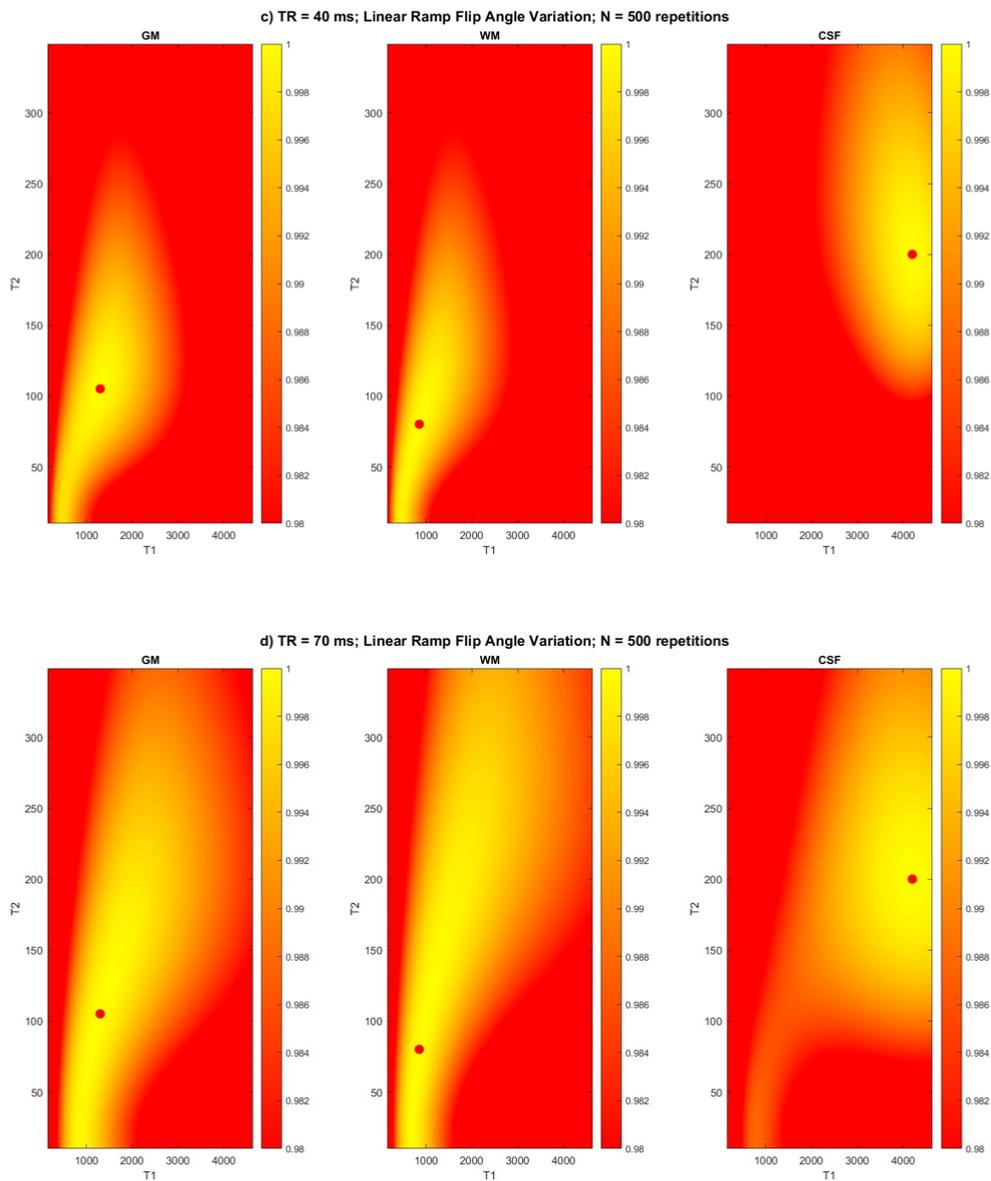

Figure 12: Supplementary Figure 1: Figure showing the T1-T2 sensitivity of the 'unbalanced SSFP' sequence with Linear Ramp Flip Angle Variation and N = 500 repetitions for a) TR = 8 ms; b) TR = 16 ms; c) TR = 40 ms and d) TR = 70 ms. Note that shorter TR's have better sensitivity.

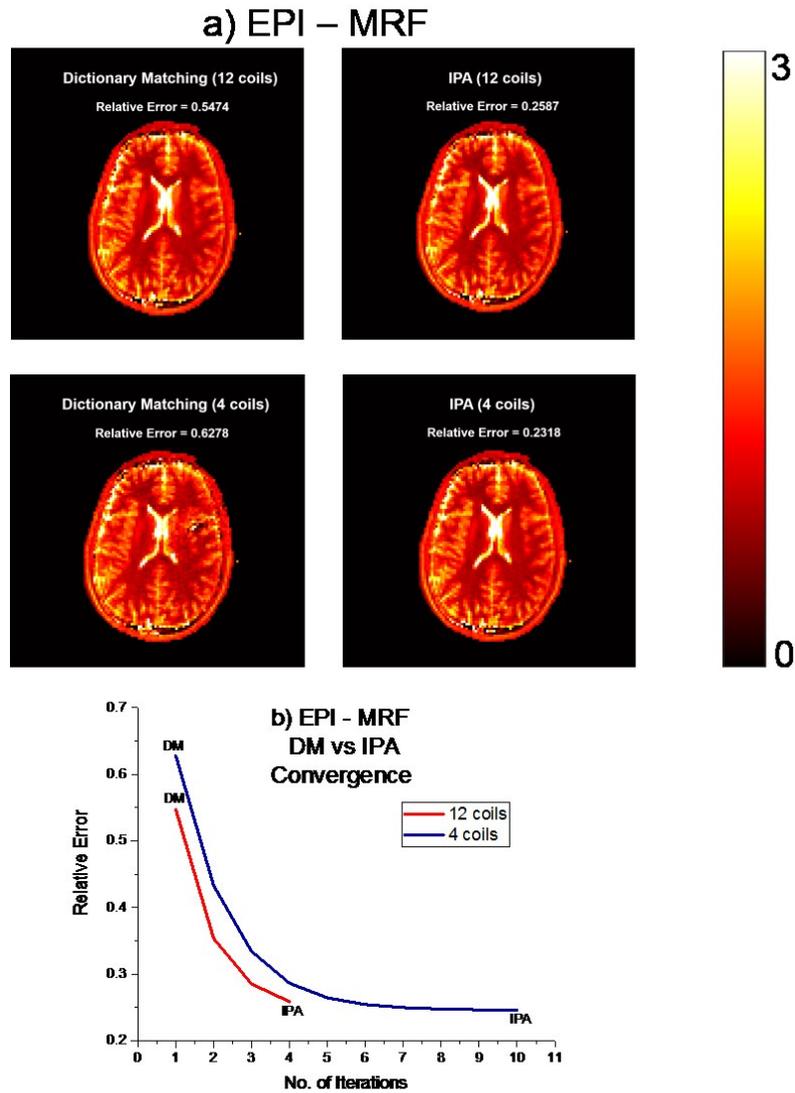

Figure 13: Supplementary Figure 2: (a) Figure showing the T1 maps of a healthy volunteer generated using 12 coils and 4 coils respectively for Dictionary Matching (DM) and Iterative Projection Algorithm (IPA) for EPI - MRF. The IPA algorithm is able to reconstruct T1 maps similar to the 12 coils case even though only 4 coils are used (this highlights the benefit of IPA in a reduced coil scenario). However, DM does not perform as well in a 4 coil scenario and produces noisy T1 maps as shown in the Figure. (b) The convergence of IPA algorithm using different number of coils are shown for EPI - MRF. Note that DM is equivalent to a single iteration of IPA.